# Cassini UVIS Observations of the Io Plasma Torus: I. Initial Results

Andrew J. Steffl

Laboratory for Atmospheric and Space Physics, 392 UCB, University of Colorado, Boulder, Colorado 80309

E-mail: steffl@colorado.edu

A. Ian F. Stewart

Laboratory for Atmospheric and Space Physics, 1234 Innovation Drive, Boulder, Colorado 80309-7814

Fran Bagenal

Laboratory for Atmospheric and Space Physics, 392 UCB, University of Colorado, Boulder, Colorado 80309

Pages: 47

Tables: 1

Figures: 13

**Proposed Running Head:** UVIS Observations of the Io torus

**Editorial correspondence to:**

Mr. Andrew J. Steffl

LASP

392 UCB

Boulder, CO 80309-0392

Phone: 303-492-3617

Fax: 303-492-6946

Email: steffl@colorado.edu

**ABSTRACT**

During the Cassini spacecraft's flyby of Jupiter (October, 2000-March, 2001), the Ultraviolet Imaging Spectrograph (UVIS) produced an extensive dataset consisting of 3,349 spectrally dispersed images of the Io plasma torus. Here we present an example of the raw data and representative EUV spectra (561Å-1181Å) of the torus, obtained on October 1, 2000 and November 14, 2000. For most of the flyby period, the entire Io torus fit within the UVIS field-of-view, enabling the measurement of the total power radiated from the torus in the extreme ultraviolet. A typical value for the total power radiated in the wavelength range of 580Å-1181Å is $1.7 \times 10^{12}$ W, with observed variations of up to 25%. Several brightening events were observed. These events lasted for roughly 20 hours, during which time the emitted power increased rapidly by ~20% before slowly returning to the pre-event level. Observed variations in the relative intensities of torus spectral features provide strong evidence for compositional changes in the torus plasma with time. Spatial profiles of the EUV emission show no evidence for a sharply peaked "ribbon" feature. The ratio of the brightness of the dusk ansa to the brightness of the dawn ansa is observed to be highly variable, with an average value of 1.30. Weak longitudinal variations in the brightness of the torus ansae were observed at the 2% level.

# 1. INTRODUCTION

The ionization of ~1 ton per second of neutral material from Io's atmosphere produces a dense (~2000 $cm^{-3}$) torus of electrons, sulfur and oxygen ions, trapped in Jupiter's strong magnetic field. While in situ, measurements of the Io plasma torus from the Voyager and Galileo spacecraft and remote sensing observations from the ground and from space-based UV telescopes have characterized the density, temperature and composition of the plasma as well as the basic spatial structure (see review by Thomas et al., 2003), the temporal variability of the torus remains poorly determined.

On its way to Saturn, the Cassini spacecraft flew past Jupiter on the dusk side of the planet with a closest approach distance of 137 $R_J$ which occurred on December 30, 2000. The optical design of the Ultraviolet Imaging Spectrograph (UVIS) enabled observations of the Jovian system from October 1, 2000 to March 31, 2001. In this paper we present an analysis of the extreme ultraviolet (EUV) emissions from the Io torus obtained by UVIS during the six-month Jupiter flyby. The 3,349 spectral images used in this analysis reveal torus variability on time scales ranging from minutes to months.

The UVIS instrument consists of two independent, but coaligned, spectrographs: one optimized for the extreme ultraviolet (EUV), covering a wavelength range of 561Å to 1181Å and the other optimized for the far ultraviolet (FUV), covering a wavelength range of 1140Å to 1913Å (McClintock *et al.* 1993, Esposito *et al.* 1998, Esposito *et al.* 2001). Each spectrograph is equipped with a CODACON 1024 x 64 pixel imaging microchannel plate detector (Lawrence and McClintock 1996). Images are obtained of UV-emitting targets with a spectral resolution of 3Å FWHM (as measured by the point-spread function) and a spatial resolution of 1 milliradian. The broad spectral range, high spectral resolution (compared to

Voyager ), and temporal coverage of UVIS resulted in the creation of a unique and rich dataset of the Io plasma torus in the ultraviolet.

In this paper, we show examples of the dataset of the Io plasma torus obtained by UVIS. Additionally, we present the temporal variability of the brightness of EUV torus emissions observed during the flyby period. In an accompanying paper, we present an analysis of observations of the midnight ansa of the Io torus made shortly after closest approach, when UVIS obtained data at highest spatial resolution.

## 2. UVIS OBSERVATIONS

The vast majority of UVIS observations of the Io torus were obtained while the Cassini spacecraft was oriented such that the spacecraft -Y axis (which points in the nominal direction of the UVIS boresight) was pointed towards Jupiter while the +X axis was pointed at the north ecliptic pole. The long axis of the UVIS entrance slit is parallel to the spacecraft Z axis and was therefore approximately parallel to Jupiter's equator. Figure 1 shows the viewing geometry for the UVIS data taken on October 1, 2000 and December 4, 2000. The viewing geometry for these observations is representative of the UVIS Io torus dataset.

[**FIGURE 1**]

For the purposes of analyzing the UVIS Io torus observations, the Jupiter flyby period can be subdivided into three separate phases, based on the location of the Cassini spacecraft and the mode of observation during that phase: the inbound staring phase, which lasted from October 1 to November 14, 2000; the inbound mosaic phase, which lasted from November 15 to December 4, 2000; and the outbound mosaic phase, which lasted from January 22, 2001 to March 31, 2001. In the period between the inbound mosaic phase and the outbound mosaic phase, the angular size of the Io torus exceeded the field-of-view of UVIS, so measurements of the total radiated power were not possible.

During the inbound staring phase, the boresight of the instrument was pointed at the center of Jupiter, where it remained for the entirety of the observation. Spectral images of the Io torus were obtained with integration times of 1000 seconds. Due to the required spacecraft downlink time and other observations requiring a different spacecraft orientation, spectra of the Io plasma torus were not acquired continuously during this period. Rather, observations were made in a cycle lasting 12 Jovian rotations (5 days). UVIS obtained data during rotations 1, 3, 5, 6, 7, 9, and 11; this cycle was repeated nine times. We analyzed a total of 1,904 spectral images of the Io torus obtained in this mode. Since data obtained during the inbound staring phase is easier to interpret than data obtained during other phases of the Jupiter flyby, much of our initial analysis has focused on this period.

On November 15, 2000, the Cassini spacecraft entered the inbound mosaic phase of the Jupiter flyby. Prior to this date, the pointing of the spacecraft had remained fixed on Jupiter for the length of a given observation. After this time, however, the spacecraft pointing was changed in order to ensure that the Orbiter Remote Sensing instruments would be able to observe the whole of Jupiter. During this inbound mosaic phase, the spacecraft pointing moved in a 2x2 mosaic pattern, centered on Jupiter, followed by a gradual north-to-south scan through the middle of the planet. The motion of the spacecraft makes the interpretation of this data significantly more difficult, as the 1000-second UVIS integrations were not synchronized with the changes in spacecraft pointing. As a result, data taken during this mode of operation are smeared along the spectral and/or the spatial directions. In addition, the motion of the spacecraft occasionally caused parts of the torus to drift in and out of the UVIS field-of-view. Therefore, we have limited our analysis of the mosaic phase data to those spectral images that contain the entire torus within the instrument field-of-view for the entire 1000-second integration period. All spectral images meeting this criterion, (roughly one out of every five UVIS images taken after November 15, 2000) were obtained while the spacecraft was executing a gradual, north-to-south scan through the center of Jupiter, resulting in a degradation of the spectral resolution of these images. Since there was no east-west motion during this scan, the spatial resolution along the length of the slit, i.e. in the radial direction, is unaffected.

The observations made during the outbound mosaic phase are very similar to those made during the inbound mosaic phase, with the exception that different mosaic patterns were used (2x2, 2x1, etc.). The same selection criterion was used to reduce the data analyzed to only those images containing the whole torus for the entire integration period, which varied from 250-1000 seconds.

**2.1. Data Reduction.**

Proper analysis of the UVIS dataset requires careful background subtraction. In addition to the expected sources of background counts from the spacecraft's radioisotope thermal generators (RTGs) and the detector electronics, there is a pinhole light-leak in the EUV channel, that allows Lyman alpha radiation from the interplanetary medium (IPM) to fall undispersed onto the detector. The FUV channel has no such light leak, and is unaffected by the EUV channel light leak. The undispersed light falling on the EUV channel detector creates a spatially non-uniform background pattern, known as the "mesa" feature, which varies in intensity depending on the pointing of the spacecraft. The mesa feature primarily affects the short wavelength end of the detector, with an intensity that slowly increases with increasing wavelength until around 950Å (column 630), where it suddenly falls off to nominal background levels. The exact location of this drop-off is row-dependent (near column 610 for row 48 and near column 660 for row 16) and is the result of the shadow of the cylindrical shade which surrounds the detector projected onto the detector surface. The effect of the mesa pattern on the quality of the Io torus data can be seen in Fig. 2, which shows a raw spectral image of the Io torus and the various background sources.

[**FIGURE 2**]

In order to determine the spatial variation of the mesa feature across the EUV detector, a composite background image was created by averaging together 75 660-second observations of the IPM. These observations are well-suited to diagnose the nature of the mesa feature since they contain only two bright spectral lines: Lyman-$\beta$ at 1025Å and $He^+$ 584Å. Since the intensity of the mesa feature varies with the spacecraft pointing, the composite background image must be scaled to the level of background present in a given observation. Inspection of the data reveals that no detectable signal from the Io torus is present at distances greater than 10 $R_J$ from Jupiter. Therefore, this region provides an estimate of the intensity of the background level. To improve the signal-to-noise ratio of the data background, a 5x5 pixel smoothing filter was passed over the image. The smoothed data image was then divided by the composite background image, after it had been likewise smoothed, to create a scaling image. Each pixel in the scaling image represents the factor by which the corresponding pixel in the composite background image must be multiplied before it is subtracted off from the data. For each column of the detector, a line was fit to the two regions of the scaling image lying at a projected radial distance of greater than 10 $R_J$. For a given column, the values of the scaling image inside of 10 $R_J$ were replaced with values interpolated from the best-fit line for that column. This produced a scaling image valid for all regions of the detector. The unsmoothed composite background image was then multiplied by the scaling image before subtracting it from the unsmoothed data image. This method effectively removes the mesa feature from the data and does not introduce any significant residuals.

During the inbound staring phase of observations, i.e. for data obtained before November 15, 2000, only the central 32 rows of the detector were read out, necessitating a slightly different method to scale the composite background image to the data image. For these data, the baseline level of background, i.e. that due to sources other than the mesa feature, was calculated by averaging together pixels longward of the mesa feature drop-off in a spectral region containing no significant emission from the Io torus. This constant value was subtracted from the data image, and a similar procedure was applied to the composite background image. After removing the constant basal background value from the two images, the data image was divided by the composite background image. Pixels in the resulting scaling image lying in wavelength regions containing no significant emissions from the torus were then averaged together to determine a constant background scaling factor. Once the scaling factor has been determined, the original composite background image, minus its baseline background value, was multiplied by the constant scaling factor. Finally, the scaled background image was subtracted from the data image (after the baseline level of background had been subtracted from the data image). For data to which both methods of background subtraction could be applied, the differences between the two methods were minimal.

## 2.2. Instrument Calibration

One of the challenges of working in the far-to-extreme ultraviolet region of the spectrum is obtaining an accurate detector calibration. For UVIS, this was accomplished with the use of two calibrated photodiodes, provided by the National Institute of Standards and Technology (NIST). These two photodiodes are the primary standards for the radiometric sensitivity calibration of UVIS. Due to laboratory constraints and the count rates of the photodiodes, it was not possible to use the NIST photodiodes directly measure the UVIS detector effective area. Rather, the NIST photodiodes were used to calibrate two secondary standards which were then used to measure the UVIS effective area. Results from the calibration standards were compared to instrumental sensitivities derived from theoretical and laboratory spectra of $H_2$, $N_2$, Ne, and Ar.

A lack of available—and suitably bright—calibration sources in the wavelength range of 580Å to 925Å hindered the absolute calibration of this region of the EUV channel. However, Ajello *et al.* (1986) present the integrated intensities of four narrow features in the spectrum of Ar between 576Å and 722Å and a fifth feature centered on 925Å. The observed intensities of these emissions relative to the feature at 925Å, which was calibrated by the secondary standard, were used to scale the instrument effective area below 900Å.

Subsequent calibration work after Cassini flew past earth in 1999 revealed that the original estimate of the EUV detector sensitivity (i.e. the lab calibration) was systematically 25% too low. The 1999 post-earth correction of the laboratory calibration was used in the analysis of all data presented in this paper and can be seen in Fig. 3. Between 900Å and 730Å, the instrument effective area rapidly declines by about a factor of three, which is consistent with pre-calibration models of the EUV channel effective area based on the reflectivity of the boron carbide coatings of the grating and the telescope (Bill McClintock, personal communication).

**[FIGURE 3]**

Work on the UVIS calibration is ongoing. Coordinated observations of the sun with UVIS and the Solar EUV Experiment (SEE) (Woods *et al.* 1998) aboard the Thermosphere, Ionosphere, Mesosphere, Energetics, and Dynamics (TIMED) spacecraft were obtained in July and December 2002. These observations will be used to cross-calibrate the EUV channel, especially in the region below 900Å, where the shape of the EUV effective area curve is poorly constrained due to paucity of laboratory calibration measurements.

## 3. RESULTS

### 3.1. UVIS Spectral Image

Spectrally dispersed images of the Io torus are presented in Fig. 4.

**[FIGURE 4]**

These images are from the EUV channel of UVIS and are typical of the quality of data obtained during the Jupiter flyby. The data were obtained during 46-hour period beginning on 11-Nov-2000 03:06 UT. Individual images, acquired when the sub-Cassini System III longitude ($\lambda_{III}$) was within ± 15° of 20°, 110°, 200°, and 290°, were averaged together to increase the signal-to-noise ratio. Jovian north is toward the left, while the dusk ansa of the torus is toward the top of the figure. The longitudes of 110° and 290° show the torus in an edge-on orientation. The Cassini spacecraft was located at 4° north latitude during these observations, and the centrifugal equator of the Io torus is inclined approximately 6° to the rotational equator. Therefore, when the Cassini spacecraft is located near $\lambda_{III} = 20°$, the opening angle of the torus is approximately 2° and the torus appears close to edge-on, while at $\lambda_{III} = 200°$, the opening angle is approximately 10°, resulting in a torus that appears somewhat face-on. A movie—created by summing the major torus emissions—showing the rotation of the Io torus during this period can be found at: http://lasp.colorado.edu/cassini/whats_new/.

Roughly 20 spectral features due to the Io plasma torus can be seen in the EUV channel image. Several of these features overlap as a result of the latitudinal extent of the Io torus and the spectral resolution of UVIS (3Å FWHM, as determined by the point-spread function). Emission from the Jovian aurora can be clearly seen in the central rows, longward of 845Å. At wavelengths below this point, Jupiter becomes an absorber rather than an emitter, and these rows appear darkened.

### 3.2. UVIS EUV Spectrum of the Io Plasma Torus

The spectrum of the dusk ansa of the Io plasma torus at the beginning of the flyby period, as observed by the EUV channel of UVIS, is shown in Fig. 5.

**[FIGURE 5]**

This spectrum is the average of a single row from each of 164 individual spectral images, obtained during a 67-hour period beginning at 08:43 UT October 1, 2000. Cassini was 1,160 $R_J$ from Jupiter at the time of these observations, so that single row covers the dusk ansa of the torus from 6.0-7.2 $R_J$ in projected radial distance. The features in the spectrum are labeled by the ion species responsible for the majority of emission in the feature and the central wavelength of the feature. Since the Io torus is clearly not a point source, the effective resolution of the spectrum is limited by the angular size of the torus in the latitudinal direction, and, as such, is proportional to the spacecraft's distance from Jupiter.

The highest resolution spectrum of the Io torus obtained by UVIS, which also includes the FUV channel, is presented in the accompanying paper (Steffl et al., 2003).

**3.3 Dawn/Dusk Brightness Asymmetry**

The dawn/dusk brightness asymmetry of the Io torus, discovered in Voyager UVS data by Sandel and Broadfoot (1982), has been seen by a variety of observers and instruments in wavelengths ranging from the ultraviolet to the infrared (Morgan 1985, Oliversen *et al.* 1991, Dessler and Sandel 1992, Hall *et al.* 1994a, Hall *et al.* 1994b, Schneider and Trauger 1995, Gladstone and Hall 1998, Herbert and Sandel 2001, Herbert *et al.* 2001, Lichtenberg *et al*. 2001). The dawn/dusk asymmetry is readily seen in the UVIS data, with the dusk ansa being, on average, 1.3 times as bright as the dawn ansa (integrated over the full EUV channel wavelength range). However, this ratio was observed to be highly variable. During the inbound staring phase (Oct. 1—Nov 14, 2000), this ratio varied from 0.73-2.29, with a mean value of 1.30 and a standard deviation of 0.25.

If there are longitudinal variations in the torus, then the instantaneous ratio of dusk ansa brightness to dawn ansa brightness will be a combination of the effects of the dawn-to-dusk electric field and the longitudinal asymmetry. The relatively short integration time ($\leq$1000 seconds), imaging capability, and temporal coverage of UVIS, make it possible to directly compare the spectrum of a particular blob of plasma located at the dawn ansa to the spectrum from the same blob of plasma (i.e. at the same longitude) when it is located at the dusk ansa, slightly less than 5 hours later. Likewise, a spectrum of the dusk ansa can be directly compared to the spectrum of the dawn ansa 5 hours later. Since we are observing the same blob of plasma as it rotates from dawn to dusk and back again, this method of obtaining a dusk-to-dawn brightness ratio will be independent of any longitudinal asymmetries.

Perhaps surprisingly, this method yields results that are virtually identical to those obtained by taking the instantaneous ratio of the two ansa brightnesses. Values of the EUV dusk/dawn brightness ratio, as observed by the Cassini UVIS, Voyager 2 Ultraviolet Spectrograph (UVS), and the Extreme Ultraviolet Explorer (EUVE), are presented in Table 1.

[TABLE 1]

Analysis of Voyager UVS data by Shemansky and Sandel (1982) determined that the dawn/dusk brightness asymmetry was caused by a 10%-30% increase in the electron temperature on the dusk side of the torus, as opposed to an increase in plasma density. Observations of the Io torus at the dawn and dusk ansae by the Hopkins Ultraviolet Telescope (HUT) indicated an electron temperature difference of ~10% (Hall *et al*., 1994a). However, the most recent analysis of EUVE observations of the Io torus determined that the dusk ansa was actually 2% cooler than the dawn ansa in 1996 and 4% cooler in 1999 (Herbert *et al.,* 2001). The uncertainty in these values is 7% for 1996 and 9% for 1999. Preliminary analyses of the UVIS spectra are consistent with an electron temperature increase of ~15% on the dusk side.

### 3.4. Radial Brightness Profiles

Each of the 1,904 spectral images obtained during the inbound staring phase of observations were summed along the spectral dimension to produce a 32 pixel-wide spatial profile of the torus EUV emissions. During the inbound staring phase, the distance of the Cassini spacecraft to Jupiter decreased by almost a factor of two, causing a corresponding factor of two increase in spatial resolution. In order to correct for errors caused by the changes in spatial scale caused by the decreasing Jovicentric distance, the individual spatial profiles were rebinned, using linear interpolation, and realigned to correct for minor pointing variations. The resulting profiles were averaged together to produce the radial profile seen in Fig 6.

[FIGURE 6]

The error bars in Fig. 6 represent the 1-σ level of the intrinsic variability of the torus radial profile, rather than the statistical uncertainty in the data. The brightness peak of both the dawn and dusk sides of the torus are located near 5.8 $R_J$. However, the spatial resolution of UVIS (~0.6 $R_J$) is insufficient to resolve the small (a few hundredths to a few tenths of an $R_J$) radial offset between the locations of the brightness peaks of the dawn and dusk ansa that is predicted by the cross-tail electric field proposed by Barbosa and Kivelson (1983) and Ip and Goertz (1983) to explain the dawn/dusk brightness asymmetry. Once scaled to the height of the dusk side, the shape of the dawn-side radial profile is nearly identical to the dusk side, outside 5.5 $R_J$. Inside of this distance, however, the dawn-side profile falls off much more slowly.

      Working independently to explain the increased temperature of the dusk ansa, as reported by Shemansky and Sandel (1982), Ip and Goertz (1983) and Barbosa and Kivelson (1983) proposed the existence of an electric field extending from dawn to dusk across the Jovian magnetosphere. Such a field might be created by the anti-sunward flow of plasma down the magnetotail. This electric field would result in higher temperatures on the dusk side via adiabatic compression of the plasma, and a radial shift in the location of the torus by a few tenths of an $R_J$ in the direction of the dawn ansa. Observations by Morgan (1985), Oliversen *et al.* (1991), Schneider and Trauger (1995), Dessler and Sandel (1992), and others have confirmed the existence of such a radial offset. Due to the relatively coarse spatial resolution of UVIS, no radial offset of the torus was detected

Radial profiles of the S III 680Å feature were also obtained by the Voyager 1 UVS (Sandel and Broadfoot 1982, Dessler and Sandel 1993) and the Extreme Ultraviolet Explorer (EUVE) spacecraft (Hall *et al.* 1994b). The radial profile of Sandel and Broadfoot (1982) was obtained from many observations during the post-encounter period. The profile of Dessler and Sandel (1993) was made from a single, high-spatial-resolution observation that also occurred during the post-encounter period. The EUVE radial profile of Hall *et al.* (1994b) was obtained during a 2 day period beginning on March 30, 1993. These profiles, along with the UVIS radial for sub-region lying with ±10Å of 680Å, are shown, after normalization to the peak of the dusk ansa, in Fig. 7.

**[Figure 7]**

The general shape of the profile observed by EUVE is fairly similar to the UVIS profile. However, the ratio of the brightness of the dawn ansa to the brightness of the dusk ansa is much higher in the EUVE profile. On the dawn side of the torus, both UVIS and UVS show a section whose brightness is nearly independent of radial distance.

The minimum energy required to produce a photon observable by the EUV channel of UVIS is 11 eV. Since the electron temperature of the torus inside 6 $R_J$ is significantly less than this (~5eV) [Sittler and Strobel, 1987], observations of this region are particularly susceptible to viewing geometry effects caused by the superposition of the warm, outer torus on the cold, inner torus. For this reason, and because the peaks in the spatial profile are asymmetrical, we believe that the half-width at half-maximum (HWHM), measured radially outward from the peak, is a more accurate measure of the intrinsic width of the torus emitting region than the full-width at half-maximum (FWHM) that has been reported in previous studies. The outward HWHM of the dusk ansa of the UVIS spatial profile is 1.3 $R_J$.

The outward HWHM of the UVS dusk ansa, as presented by *Sandel and Broadfoot* [1982], is 1.0 $R_J$, while the outward HWHM for EUVE is 1.3 $R_J$—the same as measured by UVIS. The outward HWHM for the dawn ansa is 1.4 $R_J$ for UVIS, 1.1 $R_J$ for UVS, and 1.5 for EUVE.

Analysis of high spatial resolution Voyager 1 and Voyager 2 UVS radial scans by Dessler and Sandel (1993) and Volwerk et al. (1997) suggested that roughly 80-85% of the EUV emission comes from a narrow "ribbon" feature with a FWHM of 0.22 $R_J$. Such an extremely narrow peak in the EUV emitting region of the torus is not consistent with the observations of UVIS. This result is confirmed by the highest spatial resolution observations of the torus obtained by UVIS on January 14, 2001 (Steffl *et al.*, 2003).

**3.5. Temporal Variations**

One of the greatest strengths of the UVIS dataset is that observations of the Io torus in the EUV were made regularly over a period of six months by the same instrument. This allows us to separate phenomena that are intrinsically time-variable from those that vary with either local time (i.e. the dawn/dusk asymmetry) or System III longitude.

**3.5.1. EUV Radiated Power.** Measurements of the torus luminosity provide important constraints on the energy available to the torus. For typical conditions (i.e. the densities and temperatures found by Bagenal (1994) or Steffl *et al.* (2003)) observed in the Io torus, the CHIANTI atomic physics database (Dere *et al.* 1997, Young *et al.* 2003) predicts that roughly 60% of the total power radiated by the torus is emitted in the wavelength region covered by the EUV channel of UVIS. The first measurements of the torus EUV luminosity were made by the Voyager 1 UVS. Based on these measurements, Shemansky (1980) estimated a total torus luminosity of 2.5-3.5 TW. The Voyager 2 UVS measured a total luminosity approximately twice as large (Shemansky 1987). After the Voyager flybys in

1979, no further measurements of the EUV luminosity were made until 1993, when the EUVE spacecraft made its first observations of the Io torus. EUVE covered a wavelength range of 370Å to 735Å, and the observed torus luminosity in this wavelength range was ~ 0.4 TW (Hall *et al.,* 1994b). Subsequent observations of the torus by EUVE measured the power output at 0.25-0.30 TW in 1994 (Hall *et al.,* 1995), and 0.375 TW and 0.245 TW in 1996 and 1999, respectively (Herbert *et al.,* 2001). Hall *et al.* (1995) estimated that 8-27% of the power emitted by the torus was observed by EUVE, implying that the total luminosity during the EUVE observation period was roughly 1-5 TW and consistent with measurements by Voyager.

The torus EUV luminosity, as observed by UVIS, is shown in Fig. 8.

**[FIGURE 8]**

At the beginning of the UVIS observation period, the total radiated power was approximately 2.0 TW. The long-term trend in the first half of the data is of decreasing power output with time, although there are several short-term events during which the radiated power temporarily increases. The total torus EUV luminosity falls to 1.4 TW by mid-November 2000, a decrease of more than 25% in 35 days. The increase in the scatter of the data points after November 14, 2000 is the result of the switch to the "mosaic" mode of data collection, and the subsequent degradation of the quality of the data. Before this date, however, the scatter in the observed luminosity is indicative of the intrinsic short-term variations (sometimes referred to as "twinkling") of the torus. Between consecutive 1000-second integrations, the torus EUV power output is observed to vary by ±6%. This variation is an order of magnitude greater than the statistical uncertainty in the data

The trend of decreasing luminosity with time continues until the beginning of the observation gap. This gap, extending from December 4, 2000 to January 27, 2001, results from the fact that during this period, the angular size of the torus, as seen from Cassini, is greater than the angle subtended by the widest UVIS entrance slit, thus preventing UVIS from observing the whole torus simultaneously. Observations of the torus were made during this period; however, these observations were designed to take advantage of the Cassini spacecraft's relative proximity to Jupiter to obtain higher spatial resolution images and are therefore not well-suited for determining the luminosity of the torus as a whole.

**3.5.2. EUV Luminosity Events.** Superimposed on the long-term trends and the short-term "twinkling", are several events where the torus luminosity increases significantly on timescales of a few hours. The most intense event occurred on October 6, 2000 (DOY 280). This event is shown in Fig. 9.

**[FIGURE 9]**

During this event, the torus EUV luminosity increased from 1.8 TW to 2.2 TW in just 5 hours—an increase of 22%. The torus luminosity continued to increase until the end of the UVIS observation cycle. Since the observations ended before the luminosity reached a clear maximum, it is possible that the increase was even greater. When the UVIS observations resume, eleven hours later, the power output is observed to be steadily returning to its original value of ~1.8 TW. From start to finish, this event lasted roughly 20 hours, or two Jovian rotations. This event appears to have been immediately preceded by a smaller luminosity event, whose peak intensity was also not observed.

At least five similar luminosity events were observed throughout the UVIS dataset, although none are quite as dramatic as the event on October 6, 2000. One problem associated with interpreting these events is that the duty cycle of UVIS observations—9 hours of data collection followed by a break of 11 hours—is comparable to the timescale of these brightening events. The best-sampled brightening event occurred on 2-November-2000 (DOY 307). For this event, which is shown in Fig 10, the radiated power begins at an initial

**[FIGURE 10]**

level of 1.62 TW. The morphology of this event is as follows: a potential minor brightening and decay back to the initial levels lasting ~10 hours followed by a larger brightening that peaked and was still decaying at the end of the observation period. During the main event, the luminosity steadily increases to peak value of 1.84 TW in 10.8 hours. After reaching this peak value, the radiated power decays continuously for 4.4 hours, when the observation sequence ends. When observations resume 12 hours later, the torus has returned to its pre-event level. It appears that in all of these events, the decay back to the pre-event level of radiated power occurs more slowly than the increase from the pre-event level to the peak value.

Interestingly, four of the six torus events observed are preceded by a sudden brightening of the Jovian aurora roughly 10-20 hours prior to the onset of torus brightening (Pryor *et al.* 2002, Steffl *et al.* 2002). However, given the time sampling of UVIS observations and the small number of observed events, it is difficult to make a definitive statement about the correlation of torus and auroral brightenings. As of yet, no mechanism is obvious that could correlate the brightness of the Io torus and the Jovian aurora on such relatively short timescales.

**3.5.3. Luminosity of Individual Spectral Features.** The spectral resolution of UVIS enables us to determine the contribution to the total emitted power from individual spectral features. Fig. 11 shows the luminosity of the brightest spectral feature for each of the four major ion species present in the torus.

   **[FIGURE 11]**

For clarity, the running 4-hour average of emitted power has been plotted, rather than the instantaneous power. This has the effect of smoothing out the short-term "twinkling", while preserving the longer-term trends. The S III line at 680Å is by far the most energetic feature in the EUV spectrum of the torus. This single feature, composed of some 16 individual spectral transitions, is responsible for fully 20% of the total EUV radiated power.

   Significant changes in the EUV spectrum of the torus are observed during the Jupiter flyby. At the start of the UVIS observation period ~0.10 TW of power are radiated in the S II 765Å line, while ~0.06 TW are radiated in the S IV 748Å line. After a brief jump, associated with the brightening event of October 6, 2000, the intensity of the S II line steadily decreases until it reaches a value of 0.04 TW around day 323 (November 18, 2000), where it remains until the data gap. The S IV line, on the other hand, begins near 0.06 TW—fainter than the S II line. However, whereas the intensity of the S II line decreases with time, the intensity of the S IV line initially increases with time, reaching a maximum value of 0.10 TW on day 323. After this point, it fades gradually to a value of 0.07 TW at the beginning of the data gap. When the whole-torus observations resume after closest approach, the behavior of the S II and S IV lines has settled down. The S II line intensity remains near 0.05 TW until shortly after day 60 of year 2001, after which it decreases to 0.04 TW. During this time, the S IV line exhibits variations about 0.07 TW, but shows no similar decrease after day 60 of year 2001. These trends are confirmed in the intensities of other observed spectral features resulting from S II and S IV emission. Spectra from the dusk half of the torus from the beginning and end of the inbound staring phase are presented in Fig. 12.

[FIGURE 12]

The observed changes in the torus EUV spectrum strongly suggest that significant compositional changes occurred during the UVIS observing period. For example, the ratio of S II 765Å to S IV 748Å decreased by more than a factor of three from October 1 – November 14, 2000. Since spectral features primarily result from transitions to the ground state and are close to each other in wavelength their ratio is relatively insensitive to changes in the electron temperature in the torus, i.e. if the composition of the torus remained constant, the electron temperature would have to increase by more than two orders of magnitude to account for the factor of three decrease in the brightness ratio. Such a temperature increase is clearly not physical, so the change in the brightness ratio must be caused by changes in the S II and S IV ion densities in the torus.

There are over 400 individual radiative transitions from the major ion species present in the torus that lie in the wavelength region covered by the EUV channel of UVIS, according to the CHIANTI atomic physics database. With the spectral resolution of UVIS almost all of the observed spectral features are blends of several individual radiative transitions. Therefore, it is necessary to develop a detailed model of the torus emissions in order to extract the electron temperature and ion mixing ratios from the UVIS data. Such a model, based on the CHIANTI atomic physics database has been developed and is described in Steffl *et al.* (2003). Applying this model to the full UVIS dataset will be the focus of future work.

### 3.6 System III Variations

Nearly all temporally extended observations of the Io plasma torus have looked for a correlation between torus brightness and System III longitude. Some of the most dramatic System III variations were reported by Schneider and Trauger (1995), who observed the intensity of the [S II] 6731Å feature over six nights in 1992. They reported that the longitudes 150° < $\lambda_{III}$ < 210° were consistently 3-4 times as bright as longitudes 0° < $\lambda_{III}$ < 70°. Other observers have either not detected any consistent variation with System III longitude or reported variations with lower amplitudes (Sandel and Broadfoot 1982, Morgan 1985, Brown 1995, Woodward *et al.* 1997, Herbert and Sandel 2000, Lichtenberg 2001, and references therein). Initial examination of the Cassini UVIS Io torus dataset yields a small (~5%) variation in the ansa brightness with System III longitude, over the 44 days of the UVIS inbound staring phase. This variation is shown in Fig. 13.

[FIGURE 13]

To correct for the dawn/dusk brightness asymmetry and long-term variations in torus luminosity, each individual measurement of the ansa luminosity has been divided by the ansa luminosity averaged over a 9.925-hour period, centered on the measurement time. In contrast to preciously observed System III variations, the peak in ansa brightness occurs near $\lambda_{III}$ = 110°. It should be noted that the scatter in the data is considerably larger than the amplitude of the variation, and therefore, the physical significance of the variation is uncertain.

### 4. CONCLUSIONS

The broad spectral range and high spectral resolution of UVIS combined with roughly six months of observations made during the Cassini Jupiter flyby resulted in a rich and unique dataset of EUV emissions from the Io plasma torus. In this paper we have presented initial results from the analysis of this dataset. Our main conclusions are:

1. The total power emitted by the torus in the EUV region of the spectrum (561Å-1182Å) is highly variable. Variations of 5% were observed on timescales of ~15 minutes, and longterm variations of 25% were observed on timescales of ~40 days. Intermediate-length events, lasting roughly 20 hours from start to finish, change the total torus EUV luminosity by ~20%.

2. The composition of the torus changes on timescales of a few tens of days. The observed factor of 3 variation in the ratio of S II 765Å/S IV 748Å is too great to be explained solely by a change in electron temperature.

3. The dawn/dusk brightness asymmetry of the torus ansae is highly variable. Observed values of the dusk ansa brightness divided by the dawn ansa brightness range from 0.74-2.29, with a mean value of 1.32 and a standard deviation of 0.25.

4. The profile of the brightness of the S III 680Å feature versus distance from Jupiter has a half-width at half-maximum, as measured radially outward from the peak, of 1.3 $R_J$. This width is comparable to that measured by Voyager UVS and EUVE. We find no evidence for an extremely narrow ribbon feature, such as that described by Dessler and Sandel (1993).

5. The EUV power emitted by the torus ansae over a 44 day period was observed to vary with System III longitude at the 5% level, with a peak occurring near $\lambda_{III} = 110°$. However, given the large scatter in the data points, the physical significance of this variation is uncertain.

The UVIS Io plasma torus dataset holds substantial scientific potential. Future work will include modeling the EUV emissions to derive parameters such as electron temperature and ion mixing ratios as a function of time, local time (i.e. dawn vs. dusk), and System III longitude. These parameters will provide important constraints for torus chemistry models, such as that presented by Delamere and Bagenal (2003), used to determine the plasma conditions underlying the EUV emissions.

**ACKNOWLEDGEMENTS**

Analysis of the Cassini UVIS data is supported under contract JPL 961196. FB acknowledges support as Galileo IDS under contract JPL 959550. The authors wish to thank Bill McClintock and the rest of the UVIS science and operations team for their support. CHIANTI is a collaborative project involving the NRL (USA), RAL (UK), and the Universities of Florence (Italy) and Cambridge (UK).

**Table 1.** Observed Values of the Dusk/Dawn Brightness Ratio

| Instrument | Cassini UVIS[a] | Cassini UVIS[b] | Voyager 2 UVS[c] | EUVE[d] |
|---|---|---|---|---|
| **Mean** | 1.30 | 1.31 | 1.22 | |
| **Std. dev.** | 0.25 | 0.24 | 0.33 | |
| **Min** | .73 | 0.66 | 0.64 | 1.24 |
| **Max** | 2.29 | 2.16 | 2.17 | 2.3 |
| **Percent of time Dawn > Dusk** | 10% | 7% | 22% | not observed |

[a] 44 days of observations from Cassini inbound staring phase in 2000 Instantaneous ratio of dusk ansa brightness / dawn ansa brightness

[b] Dusk ansa brightness / dawn ansa brightness (1/2 rotation later) combined with Dusk ansa brightness (1/2 rotation later) / dawn ansa brightness

[c] 8 days of observations from Voyager 2 inbound leg in 1979 (Sandel and Broadfoot, 1982)

[d] 2 days of observation in 1993 (Hall *et al.*, 1994), 5 days in 1996 (Gladstone and Hall, 1998), and 15 days in 1999 (Herbert *et al.*, 2001)

**FIGURE CAPTIONS**

**Figure 1**. Viewing geometry for UVIS observations of the Io torus on October 4, 2000 and December 4, 2000. The projection of the UVIS occultation slit is shown relative to the locations of the Jupiter and the Galilean satellites and their orbits.

**Figure 2**. Unprocessed image of the Io torus. This image illustrates the signal-to-noise of the UVIS data and the various background effects. The dusk ansa of the torus is at the top half of the image, and Jovian north is to the left. The background level in the "mesa" feature is a factor of ~2 higher than the nominal background. The faint, diamond-shaped pattern present in the mesa feature is caused by the shadow of a wire grid positioned above the detector. Scattered Lyman alpha can also be seen at the long wavelength end (right) of the EUV detector.

**Figure 3.** Effective area (cm$^2$) of the EUV and FUV channels of UVIS. The error bars represent the 1-σ error from the quadrature sum of the errors from the NIST diode formal error, the transfer of calibration to a secondary calibration standard, and the measurement precision. Figure from Bill McClintock (personal communication).

**Figure 4.** UVIS EUV channel spectral images showing the Io torus in open and edge-on configurations. As in Fig. 2, the dusk ansa is up, and Jovian north is to the left. The central meridian longitude (CML) of the Cassini spacecraft during the observation is given to the left of the spectral images. To increase signal-to-noise, observations obtained over a 46-hour period beginning on 11-Nov-2000 03:06 UT with a CML ±15° of the nominal value were averaged together. Cassini was located at 4° north latitude during the observations, causing the image taken at CML $\lambda_{III}$=20° to appear near edge-on. Auroral emissions from Jupiter are visible in the central rows. Shortward of ~845Å, Jupiter becomes absorber. The narrow, vertical features near 600Å are instrumental artifacts.

**Figure 5.** Composite EUV (561Å-1181Å) spectrum of the Io plasma torus dusk ansa. This spectrum is the average of 164 1000-second integrations obtained during a 67-hour period beginning at 08:43 UT October 1, 2000. Cassini was ~1,160 $R_J$ from Jupiter at the time of these observations, so the spectrum was extracted from 6.0-7.2 $R_J$ in projected radial

distance. The spectral features are labeled by wavelength and the ion species that makes the dominant contribution to the feature.

**Figure 6.** Total UVIS EUV channel (561Å-1181Å) spatial profile. The dawn ansa is to the left while the dusk ansa is to the right. The error bars represent the intrinsic 1-σ variance of the Io torus i.e. the instantaneous observed spatial profile lies within the error bars 68% of the time.

**Figure 7.** Spatial profiles of the S III 680Å feature. The thick black line is the UVIS profile, averaged over 41 days. The connected grey dots are the Voyager 1 UVS profile of Sandel and Broadfoot (1982). The thin, smooth, black line is the high-resolution UVS profile of Dessler and Sandel (1993). The histogram is the EUVE profile of Hall *et al.* (1994). All three instruments observe the brightness of the dawn side to fall off more slowly inside of the peak on the dawn side than on the dusk side.

**Figure 8.** Total torus EUV luminosity during the UVIS observation period. The vertical spread of the data points before November 14, 2000 is due to the intrinsic variability of the torus. After this date, the spread is a combination of intrinsic variability and uncertainty introduced by the "mosaic" mode of UVIS operation. The observation gap in the middle of the plot is a result of angular size of the torus exceeding the UVIS occultation slit field of view.

**Figure 9.** Total torus EUV luminosity versus time for the Day 280 event. The total torus luminosity increases sharply around DOY 280.6. The decay back to "normal" torus levels is less rapid. The observed scatter in the data points greatly exceeds the statistical noise in the data and is intrinsic to the torus (i.e. twinkling).

**Figure 10.** Total torus EUV luminosity versus time for the Day 307 event.

**Figure 11.** Total power radiated in four spectral features. The thick line is the power emitted in the S IV 748Å feature. The data have been averaged over one Jovian rotation period. Changes in the S II 765Å to S IV 748Å ratio are evidence of compositional changes in the Io torus.

**Figure 12.** EUV spectra from the dusk ansa of the torus on October 1 and November 14, 2000. Since the effective spectral resolution is determined by the angular size of the torus, the resolution decreases as Cassini approaches Jupiter. Note the dramatic change in the S II 765Å to S IV 748Å ratio.

**Figure 13.** Relative EUV luminosity of the torus ansae versus System III longitude. Both dawn and dusk ansae are included in this plot. The thick line is the average of all data points in 10° longitude bins.

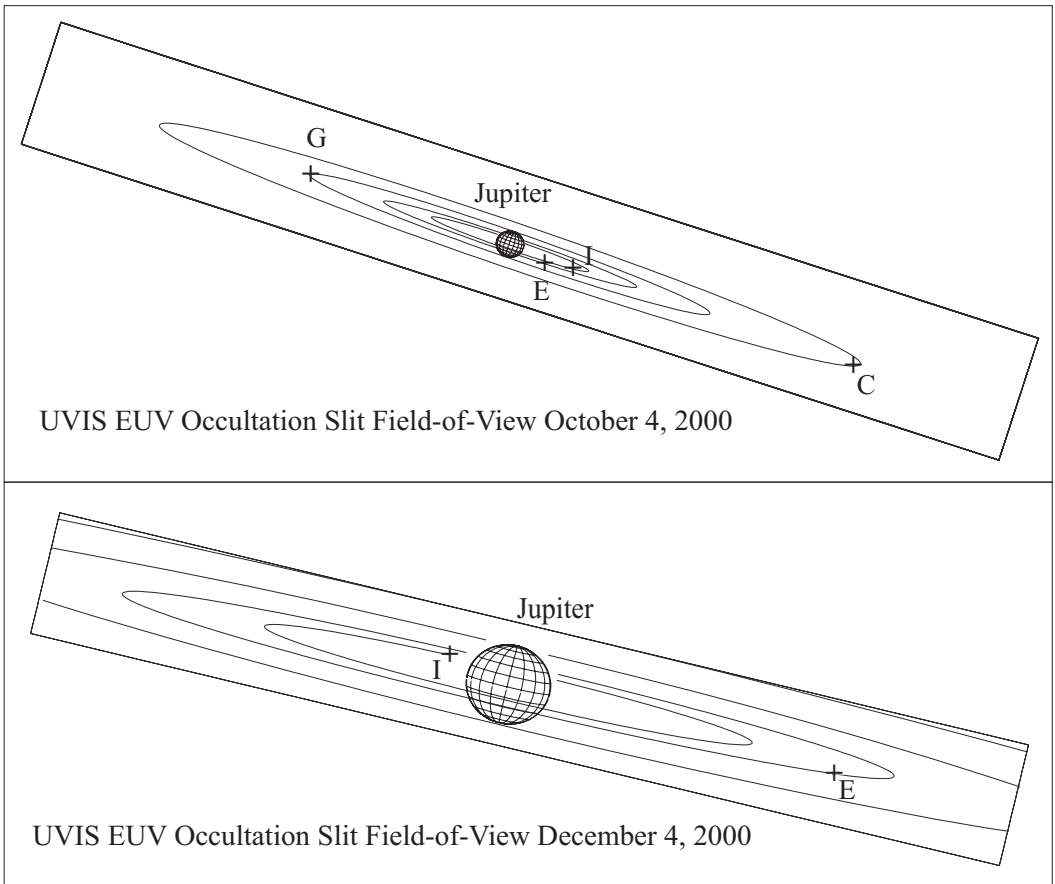

FIGURE 1. Steffl *et al.*, UVIS Observations of the Io torus

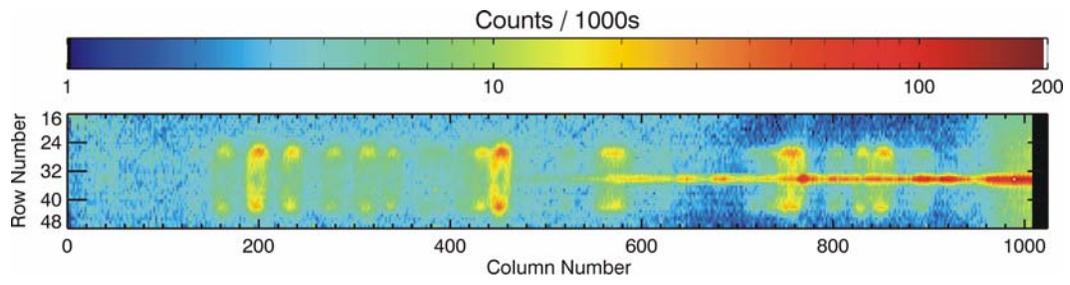

FIGURE 2  Steffl *et al.*, UVIS Observations of the Io torus

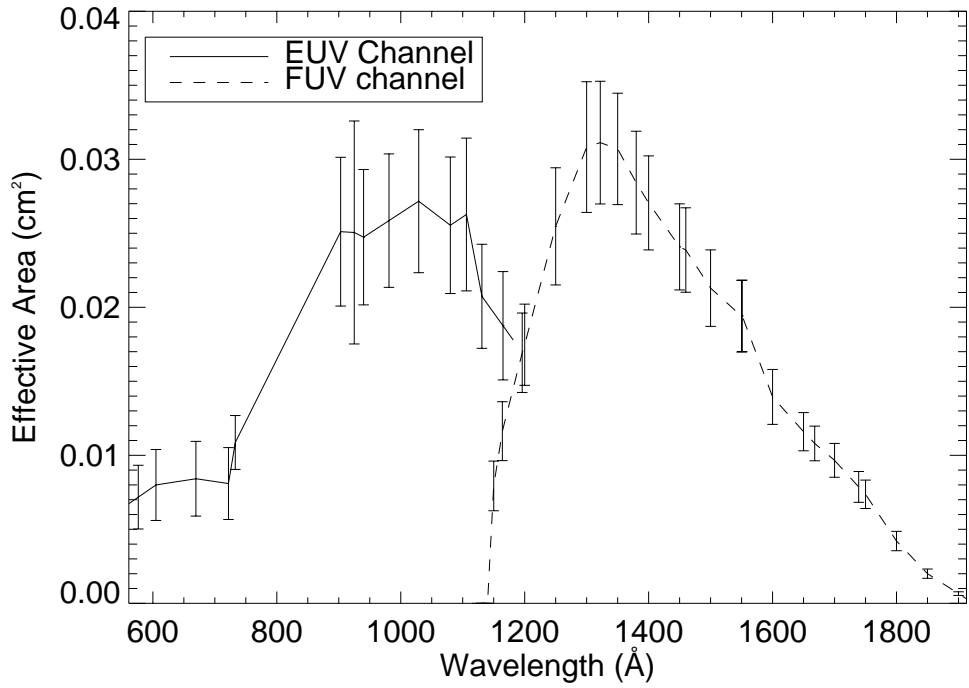

FIGURE 3. Steffl *et al.*, UVIS Observations of the Io torus

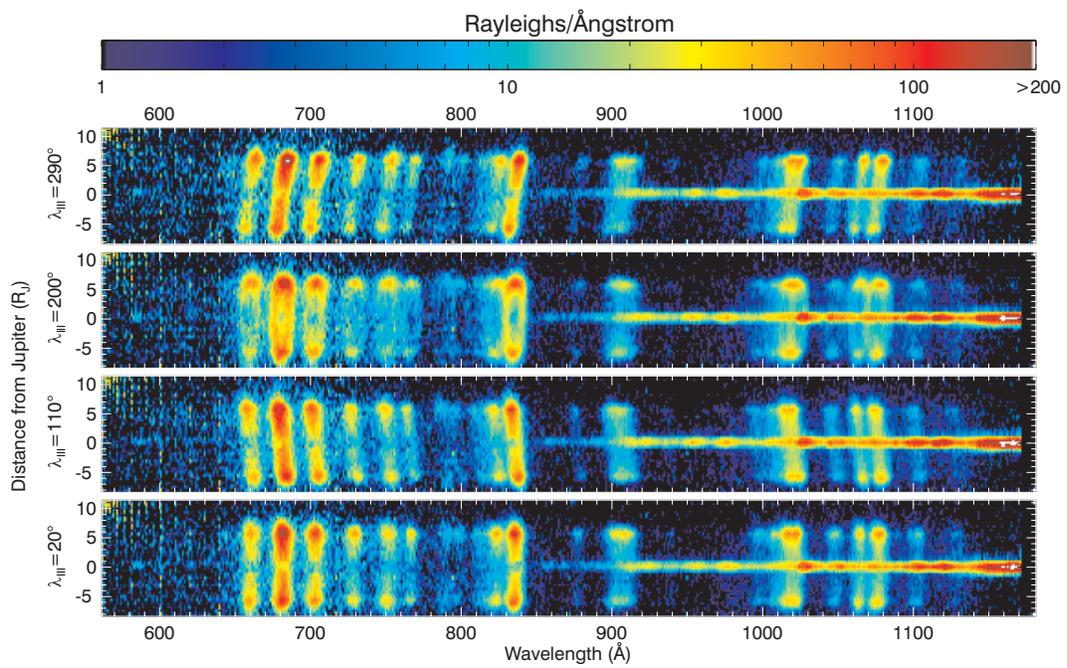
FIGURE 4. Steffl *et al.*, UVIS Observations of the Io torus

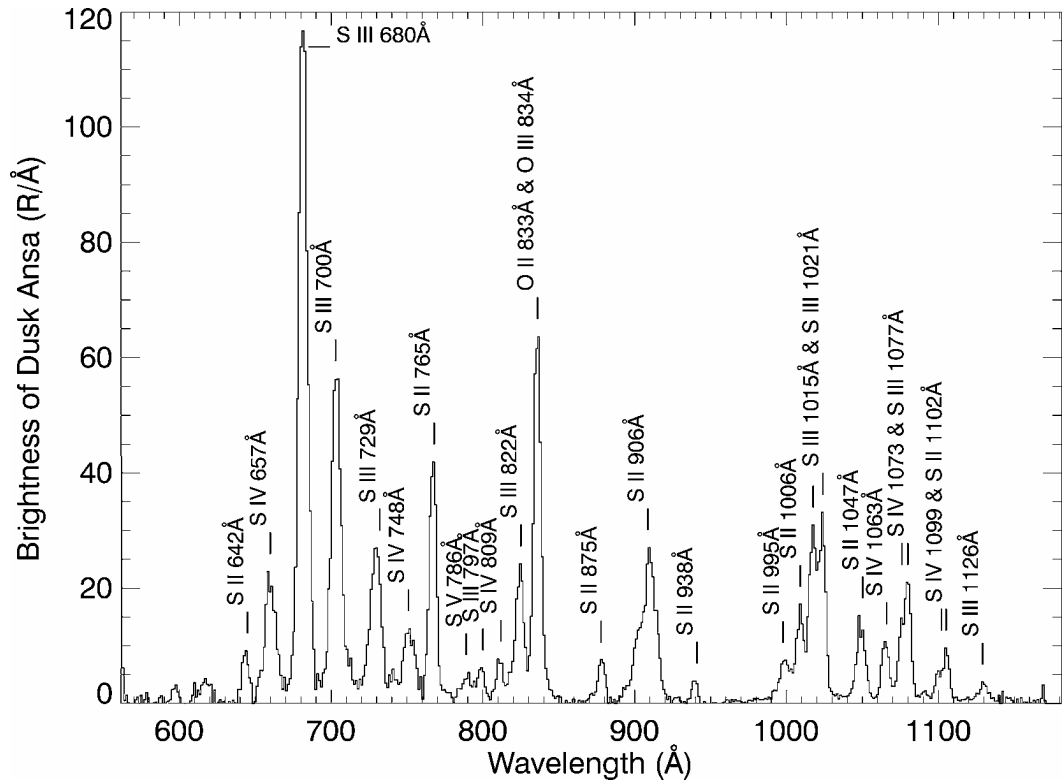
FIGURE 5. Steffl *et al.*, UVIS Observations of the Io torus

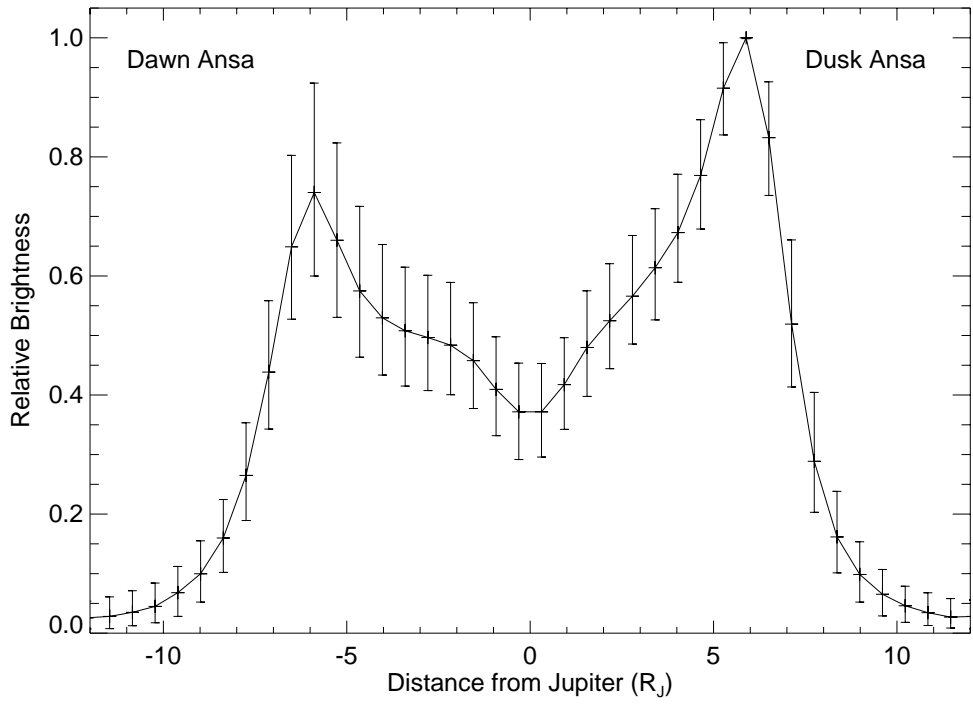

FIGURE 6. Steffl *et al.*, UVIS Observations of the Io torus

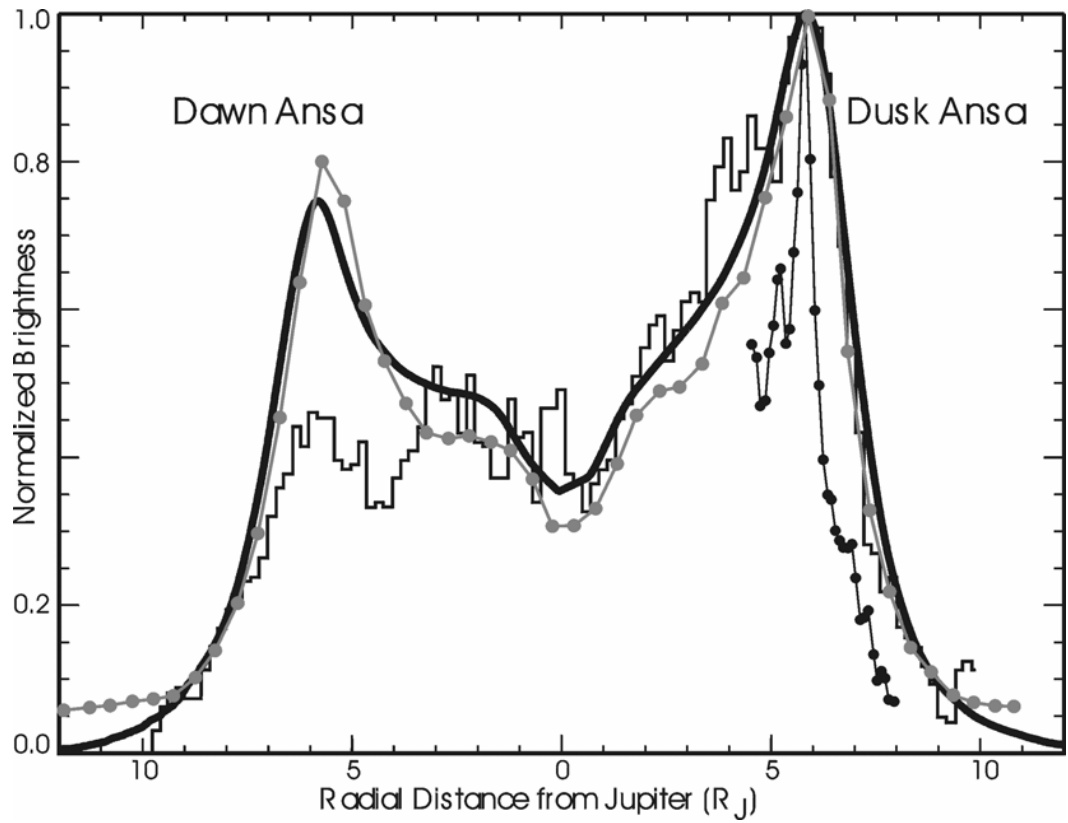

FIGURE 7. Steffl *et al.*, UVIS Observations of the Io torus

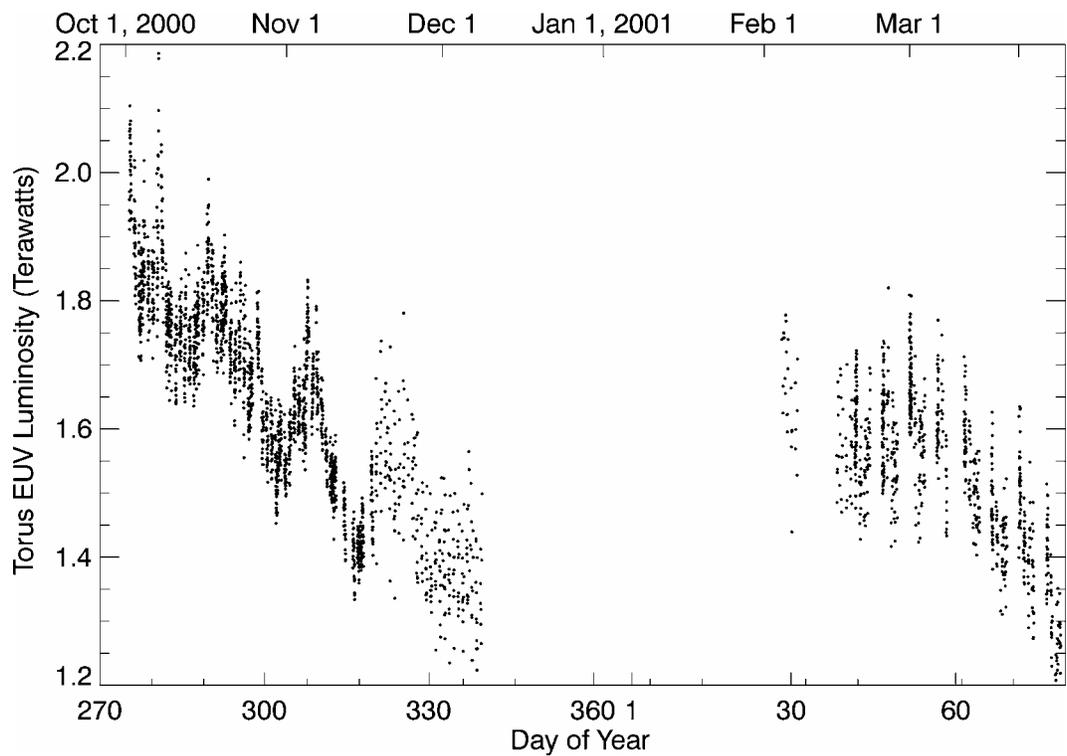
FIGURE 8. Steffl *et al.*, UVIS Observations of the Io torus

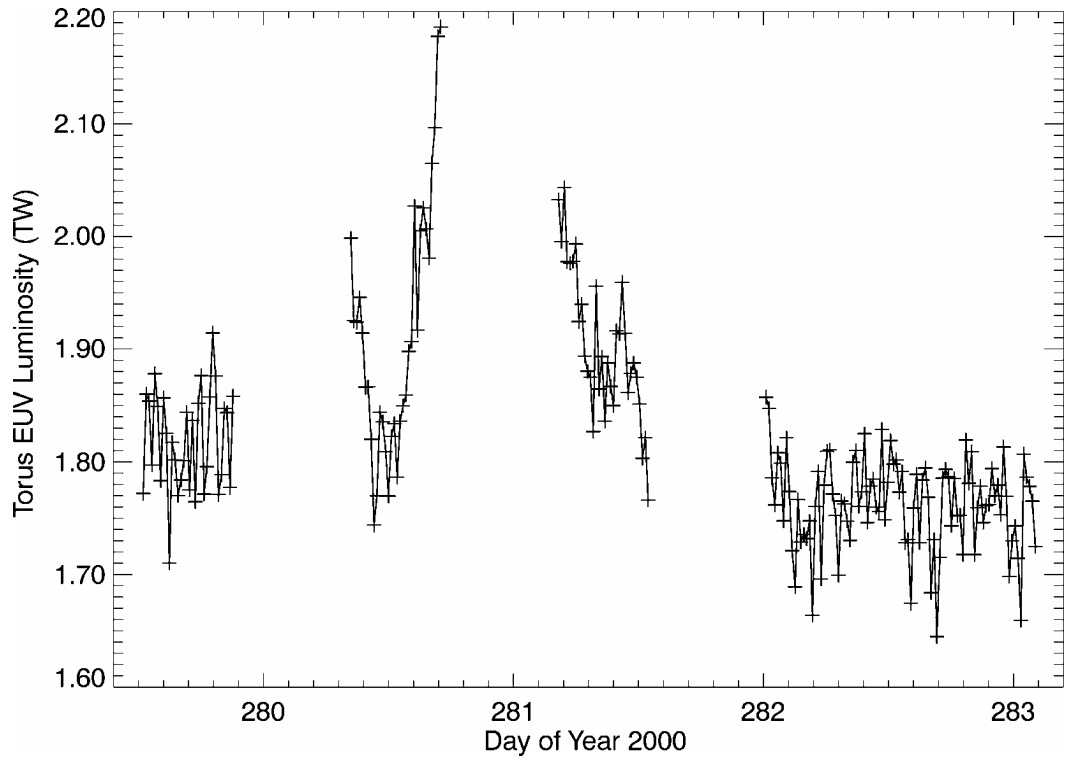
FIGURE 9. Steffl *et al*, UVIS observations of the Io torus

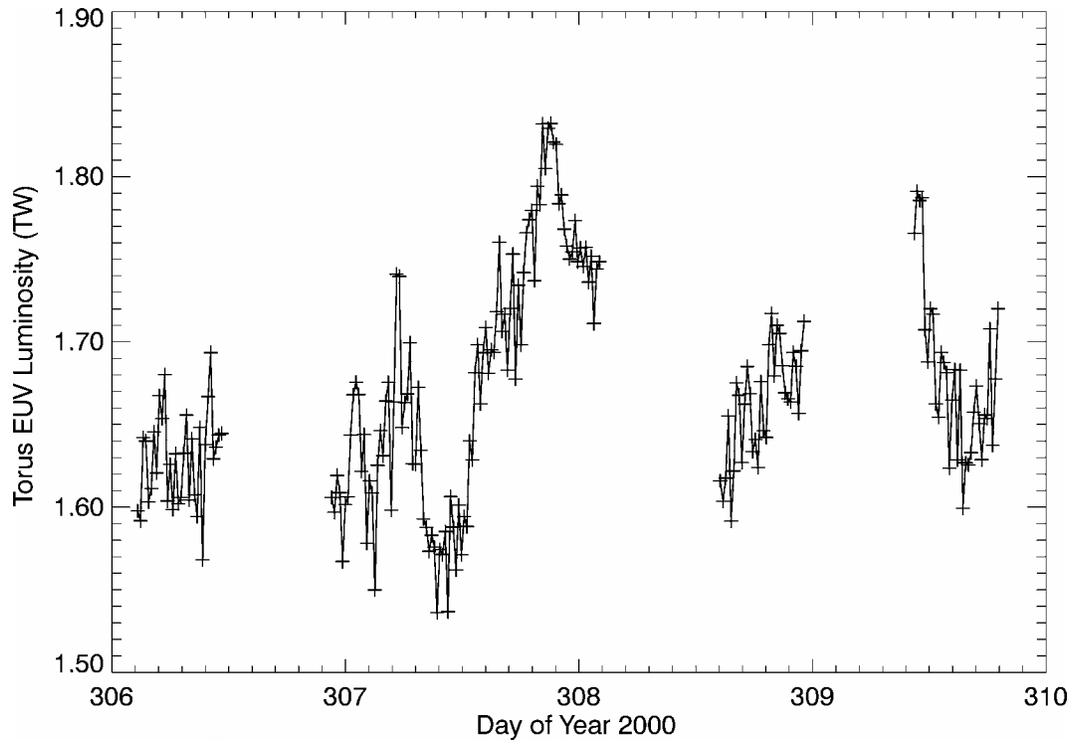
FIGURE 10. Steffl *et al*, UVIS observations of the Io torus

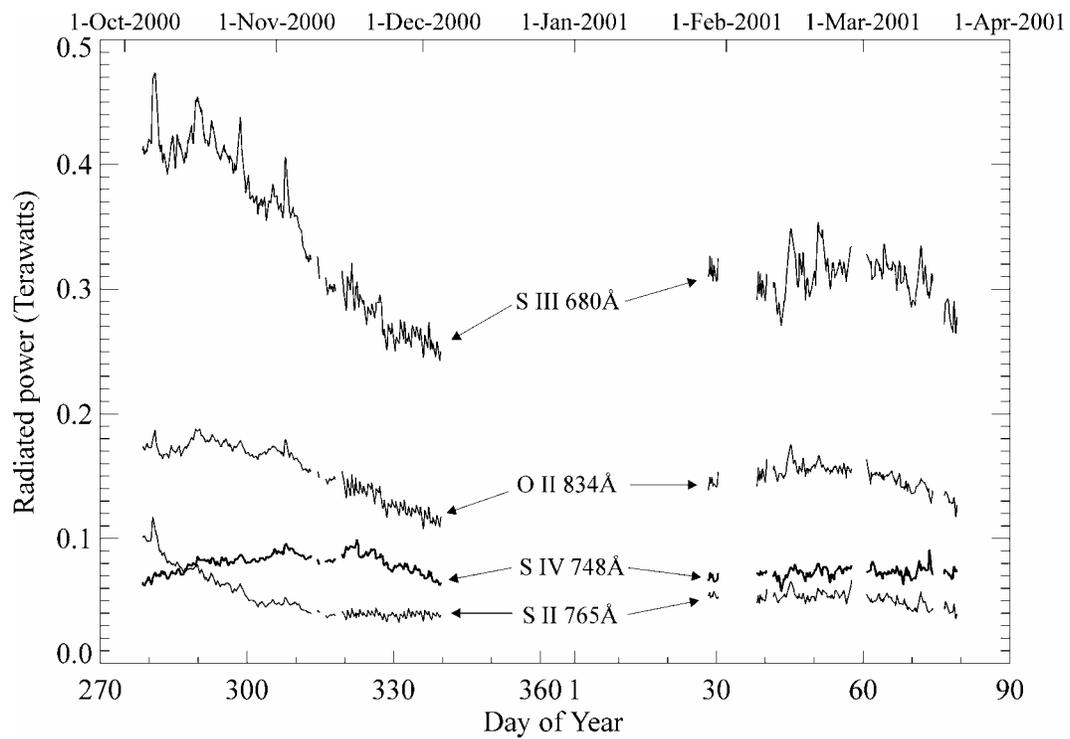
FIGURE 11. Steffl *et al*., UVIS Observations of the Io torus

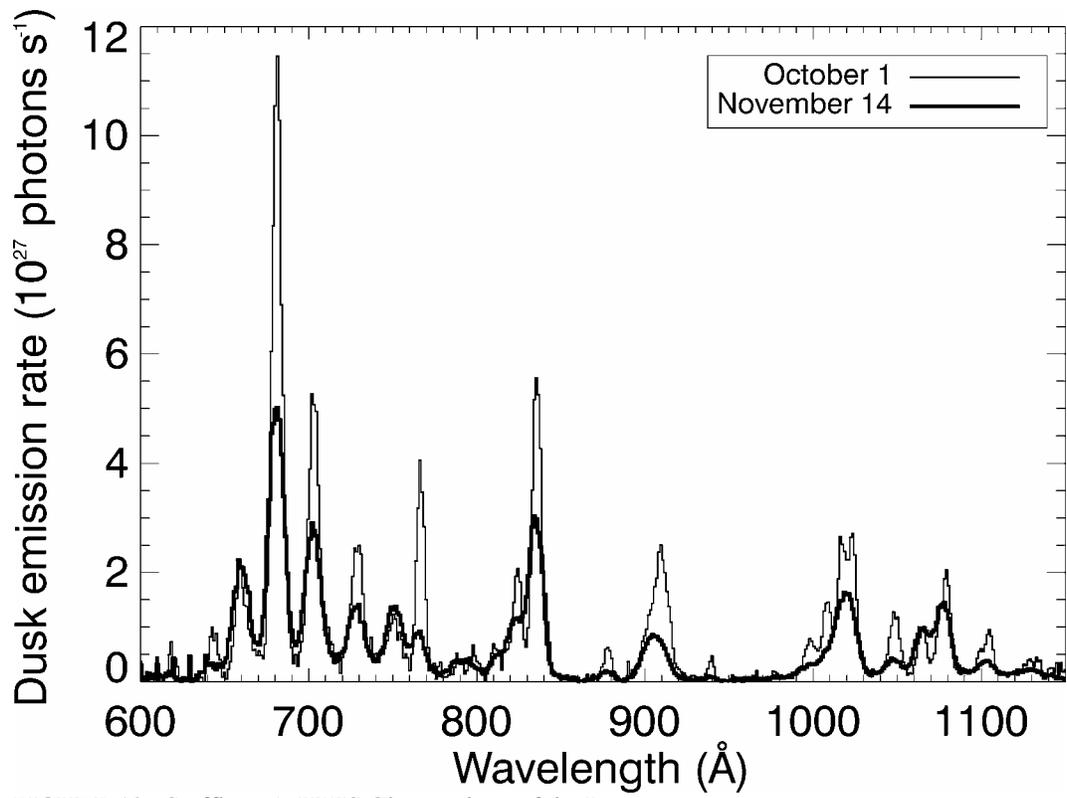
FIGURE 12   Steffl *et al.*, UVIS Observations of the Io torus

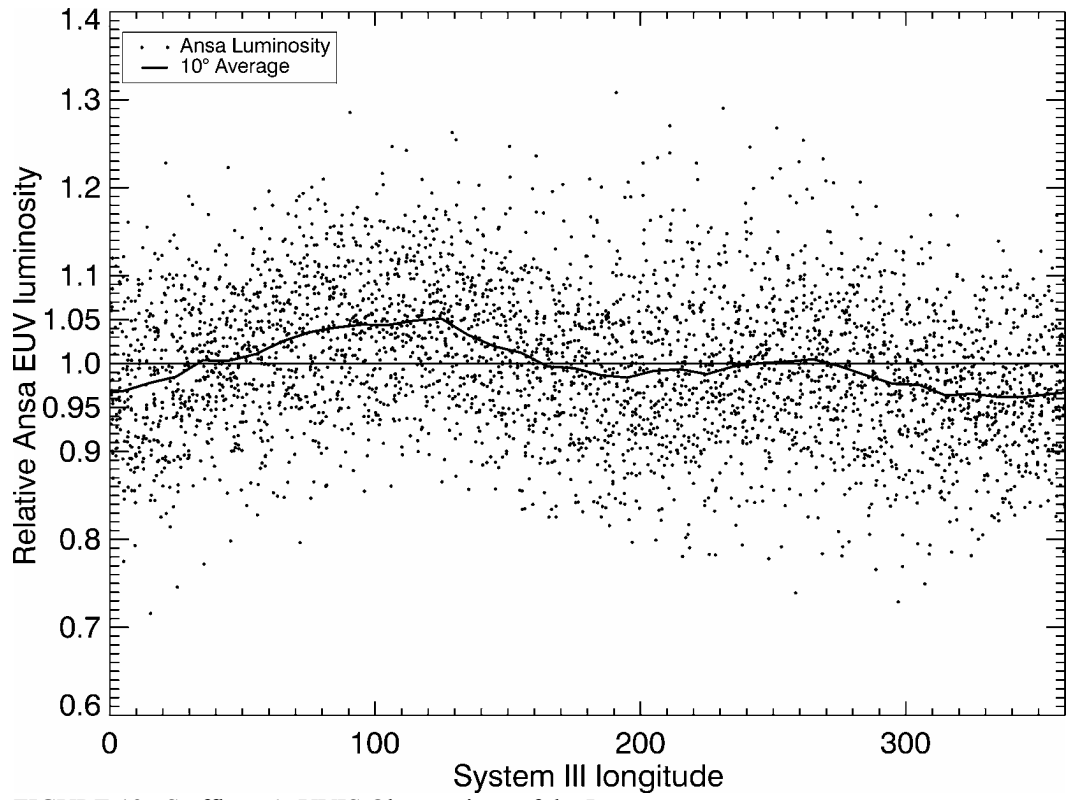
FIGURE 13   Steffl *et al.,* UVIS Observations of the Io torus